\documentclass[useAMS,usenatbib,usegraphicx]{mn2e}

\def\apj{ApJ}
\def\apjl{ApJL}
\def\aap{A\&A}
\def\mnras{MNRAS}

\def\prd{Phys.~Rev.~D}
\def\pasp{PASP}
\def\aj{AJ}
\title[A TTV analysis of seven RISE light curves of HAT-P-3.]
{A Transit Timing analysis of seven RISE light curves of the exoplanet system HAT-P-3.}

\author[N. P. Gibson et al.]{
N. P. Gibson$^{1}$\thanks{E-mail: ngibson07@qub.ac.uk},
D. L. Pollacco$^{1}$,
S. Barros$^{1}$,
C. Benn$^{2}$,
D. Christian$^{3}$,
M. Hrudkov\'a$^{4}$,\newauthor
Y. C. Joshi$^{1,5}$,
F. P. Keenan$^{1}$,
E. K. Simpson$^{1}$,
I. Skillen$^{2}$,
I. A. Steele$^{6}$.
and I. Todd$^{1}$.\\
$^{1}$Astrophysics Research Centre, School of Mathematics \&\ Physics, Queen's University, University Road, Belfast, BT7 1NN, UK\\
$^{2}$Isaac Newton Group of Telescopes, Apartado de Correos 321, E-38700 Santa Cruz de la Palma, Tenerife, Spain\\
$^{3}$Physics \& Astronomy Department, California State University Northridge, Northridge, California 91330-8268, USA\\
$^{4}$Astronomical Institute, Charles University Prague, V Holesovickach 2, CZ-180 00 Praha, Czech Republic\\
$^{5}$Aryabhatta Research Institute of Observational Sciences, Manora Peak, Nainital-263129, India \\
$^{6}$Astrophysics Research Institute, Liverpool John Moores University, CH61 4UA, UK
}

\begin{document}

\date{Accepted 2009 July. Received 2009 July; in original form 2009 July}

\pagerange{\pageref{firstpage}--\pageref{lastpage}} \pubyear{2002}

\maketitle

\label{firstpage}

\begin{abstract}
We present seven light curves of the exoplanet system HAT-P-3, taken as part of a transit timing program using the RISE instrument on the Liverpool Telescope. The light curves are analysed using a Markov-Chain Monte-Carlo algorithm to update the parameters of the system. The inclination is found to be $i = 86.75^{+0.22}_{-0.21}\deg$, the planet-star radius ratio to be $R_p/R_{\star}= 0.1098^{+0.0010}_{-0.0012}$, and the stellar radius to be $R_\star = 0.834^{+0.018}_{-0.026} R_\odot$, consistent with previous results but with a significant improvement in the precision. Central transit times and uncertainties for each light curve are also determined, and a residual permutation algorithm used as an independent check on the errors. The transit times are found to be consistent with a linear ephemeris, and a new ephemeris is calculated as $T_c(0) = 2454856.70118 \pm 0.00018$ HJD and $P = 2.899738 \pm 0.000007$ days. Model timing residuals are fitted to the measured timing residuals to place upper mass limits for a hypothetical perturbing planet as a function of the period ratio. These show that we have probed for planets with masses as low as 0.33 $M_{\oplus}$ and 1.81 $M_{\oplus}$ in the interior and exterior 2:1 resonances, respectively, assuming the planets are initially in circular orbits.
\end{abstract}

%________________________________________________________________

\section{Introduction}

Transiting planets are vital to our understanding of the structure of planetary systems, as they allow measurement of the radius of the planet, and the mass of the planet unambiguously when coupled with radial velocity data. This allows the density and surface gravity to be derived, and the internal structure of the planet may be inferred. On-going transit surveys such as SuperWASP \citep{pollacco_2008}, HATNet \citep{bakos_2002} and XO \citep{mccullough_2005} are now pushing the number of known transiting planets towards 60\footnote{see http://exoplanet.eu/catalog.php}. Most of the transiting planets are hot Jupiters, which produce a $\sim$1\% dip in the flux when transiting their host stars. However, an Earth-sized planet will produce a dip of only $\sim0.01$\%, and expensive space-based transit surveys (e.g. CoRoT, Kepler) are required to reach this level of accuracy.

Another method to detect Earth-sized planets is through measuring Transit Timing Variations (TTV). A transiting planet will maintain a constant period whilst orbiting its host star, excluding tidal effects and general relativity. Therefore, if we measure the time of mid-transit for a given system, and see variations from a constant period, we may conclude that there must be a third body in the system perturbing the orbit of the known transiting body \citep{miralda_2002, holman_murray_2005, agol_2005, heyl_gladman_2007}. TTV is particularly sensitive when the third body is in a resonant orbit, in which case sub-Earth mass planets may be detected. It is also sensitive to exomoons \citep{kipping_2009} and Trojans \citep{ford_gaudi_2006,ford_holman_2007}, and therefore has the potential to provide the first detection of an Earth-mass body orbiting a main-sequence star other than our own. 

Detecting Earth-mass bodies in resonant orbits requires the central transit times to be measured to an accuracy of  $\sim$10 seconds, for which we need high precision light curves obtained at high cadence. In theory, we can measure transit times to several seconds with a moderate-sized telescope, but in practice we are limited by correlated noise in the light curves, which arises from unknown changes in either the CCD sensitivity, telescope optics or observing conditions \citep{pont_2006}. There may also be real brightness variations in the flux of the target or comparison stars. When these sources of correlated noise are minimised, it is  possible to measure central transit times to $<10$ seconds \citep[see][for examples using space and ground-based photometry, respectively]{pont_2007,winn_holman_2009}. This allows the detection of Earth-mass planets in low-order mean-motion resonance, or more massive planets out of resonance \citep[see e.g.][]{steffen_agol_2005, agol_steffen_2007, bean_2009}.

We developed the RISE (Rapid Imager to Search for Exoplanets) instrument to obtain high precision light curves of exoplanets for TTV measurements. RISE is a fast readout camera mounted on the 2.0-m Liverpool Telescope (LT) on La Palma. It was commissioned in 2008 February, and since then there have been on going observations to observe exoplanet transits and detect a TTV signal. First results from RISE have been presented in \citet[][hereafter G08]{gibson_2008} for WASP-3, and in \citet[][hereafter G09]{gibson_2009} for TrES-3, where $\sim$10 seconds timing accuracy is achieved for the best transits.

HAT-P-3 was discovered by \citet[][hereafter T07]{torres_2007}, and is a $\sim$0.6 $M_{Jup}$ mass planet orbiting a K-type dwarf star with a period of $\sim2.9$ days. The best light curve was taken with the 1.2 m telescope at FLWO, and since then there have been no published high quality light curves of this system. Here, we present a further seven RISE transit light curves of HAT-P-3, and use them to improve the system parameters. Additionally, we extract the TTV signal in an effort to detect a third body in the system.

In Section~\ref{sect:observations} we describe the observations and data reduction, and in Section~\ref{sect:modelling} describe how the light curves are modelled and in particular how the central transit times and uncertainties are found. Our results are presented in Section~\ref{sect:results}, where we use the transit timing residuals to place upper mass limits on a perturbing planet that could be present in the HAT-P-3 system without being detected from our observations. Finally, in Section~\ref{sect:summary} we summarise and discuss our results.

%________________________________________________________________
\section{Observations and data reduction}
\label{sect:observations}

%________________________________________________________________
\subsection{RISE photometry}
\label{sect:photometry}

Five full and two partial transits were observed using the LT and RISE from 2009 January 24 to 2009 May 26. The RISE instrument is described in detail in \citet{steele_2008} and G08. It consists of a frame transfer CCD with a relatively large field-of-view (9.4 $\times$ 9.4 arcmin$^2$), and a single wide band filter ($\sim$ 500--700 nm).

For all observations, the instrument was in $2 \times 2$ binning mode, giving a scale of 1.1 arcsec pixel$^{-1}$. For the first two transits, exposure times of 5 seconds were used with the telescope slightly defocused to reduce flat fielding errors and avoid saturating the CCD. A total of 2\,520 images were obtained over a 3.5 hour period, allowing $\sim$35--45 mins both before and after the transit. For the remaining transits, exposure times of 4 seconds were used, and the telescope had slightly less defocusing, to allow 3\,150 images to be taken over a $\sim$3.5 hour period for the full transits. Due to observing constraints and weather, less than 3\,150 images were taken on most nights, and the actual number of images obtained each night is shown in Table~\ref{tab:hatp3_lightcurves}. The images have a typical FWHM of $\sim$2-4 pixels ($\sim$2.2-4.4\,arcsec), and the nights were clear for the majority of the observations, except for the night of 2009 April 24, where large scatter due to clouds can be seen in the light curve.

\begin{figure}
\includegraphics[width=84mm]{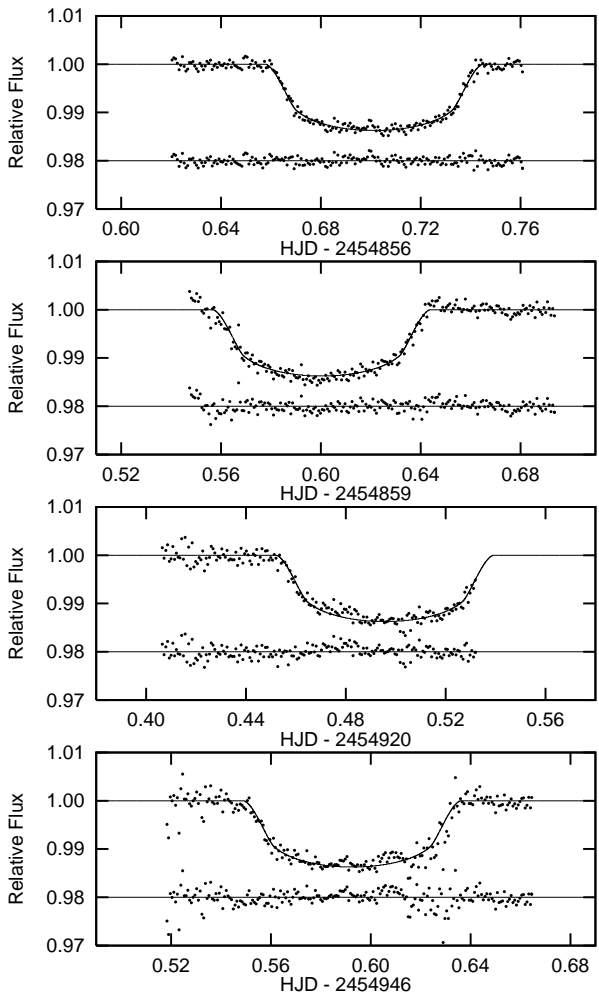}
\caption{
RISE light curves of HAT-P-3 taken from 2009 January 24 to 2009 April 24 in 1 minute bins. Their best fit models are over-plotted and the residuals from the best fit are shown offset below each light curve.
}
\label{fig:hatp3_lightcurves_a}
\end{figure}

\begin{figure}
\includegraphics[width=84mm]{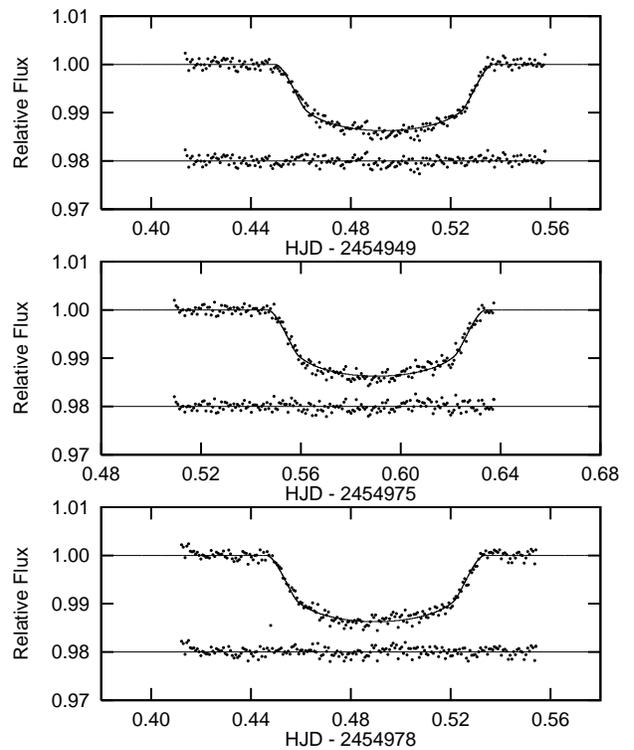}
\caption{
Same as Figure~\ref{fig:hatp3_lightcurves_a}, for the light curves taken from 2009 April 27 to 2009 May 26.
}
\label{fig:hatp3_lightcurves_b}
\end{figure}

\begin{figure}
\includegraphics[width=84mm]{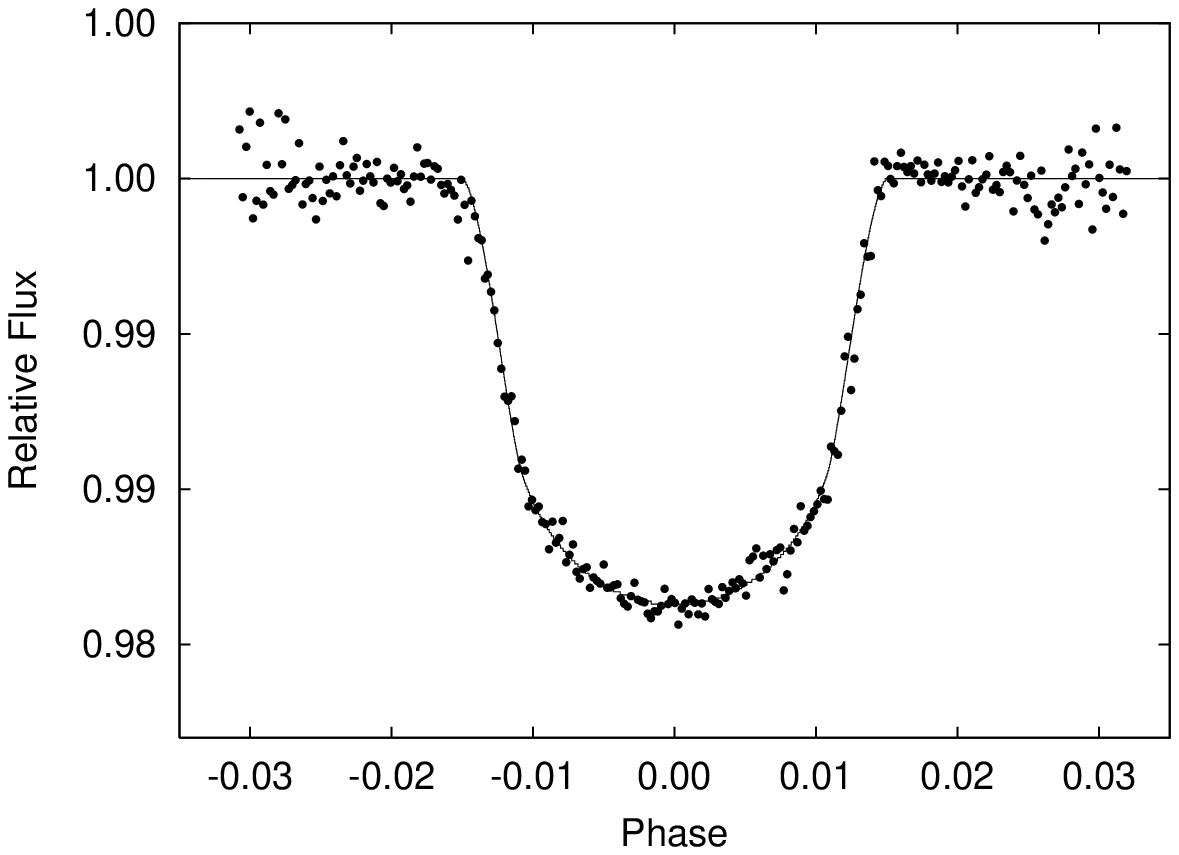}
\caption{
Phase folded light curve of the seven RISE light curves shown in 1 minute bins.
}
\label{fig:hatp3_phased_lightcurve}
\end{figure}

Images were first de-biased and flat-fielded with combined twilight flats using standard IRAF\footnote{IRAF is distributed by the National Optical Astronomy Observatories, which are operated by the Association of Universities for Research in Astronomy, Inc., under cooperative agreement with the National Science Foundation.} routines. Aperture photometry was then performed on the target and comparison stars with Pyraf\footnote{Pyraf is a product of the Space Telescope Science Institute, which is operated by AURA for NASA.} and the DAOPHOT package. The number of comparison stars and aperture size for each of the nights is shown in Table~\ref{tab:hatp3_lightcurves}. These varied as the conditions and field orientation changed for each night of observations, and were selected to minimise the out-of-transit rms. The light curves were then extracted by dividing the flux from HAT-P-3 from the sum of the flux from the comparison stars (all checked to be non-variable). Initial estimates of the photometric errors were calculated using the aperture electron flux, sky and read noise. The light curves were then normalised by dividing through with a linear function of time fitted to the out-of-transit data, setting the unocculted flux of HAT-P-3 equal to one. The normalisation parameters are allowed to vary freely when fitting the light curves, to account for any errors resulting from this procedure. The light curves, along with their best fit models and residuals (see Section~\ref{sect:system_parameters}), are shown in Figures~\ref{fig:hatp3_lightcurves_a} and  \ref{fig:hatp3_lightcurves_b} in 1 minute bins. A phase folded light curve generated from the seven RISE light curves is shown in Figure~\ref{fig:hatp3_phased_lightcurve}, again in 1 minute bins.

\begin{table*}
\begin{minipage}{110mm}
\caption{Summary of the RISE light curves of HAT-P-3, showing the number of comparison stars and aperture size used, and the rms of the residuals after placing in 1 minute bins.}
\label{tab:hatp3_lightcurves}
\begin{tabular}{lcccc}
\hline
Night & Exposures & No. comparison & Aperture size & rms (residuals)\\
~ & ~ & stars & (pixels) & (mmag)\\
\hline
2009 Jan 24 & 2520 $\times$ 5s & 5 & 7 & 0.88 \\
2009 Jan 27 & 2520 $\times$ 5s & 5 & 8 & 1.27 \\
2009 Mar 29 & 2700 $\times$ 4s & 2 & 8 & 1.37 \\
2009 Apr 24 & 3140 $\times$ 4s & 2 & 5 & 1.99 \\
2009 Apr 27 & 3150 $\times$ 4s & 3 & 7 & 0.96 \\
2009 May 23 & 2851 $\times$ 4s & 3 & 8 & 1.01\\
2009 May 26 & 3150 $\times$ 4s & 2 & 8 & 1.44 \\
\hline
\end{tabular}
\end{minipage}
\end{table*}

%________________________________________________________________
\section{Light curve modelling and analysis}
\label{sect:modelling}

%________________________________________________________________
\subsection{Determination of system parameters}
\label{sect:system_parameters}

In order to determine the system parameters from the transit light curves, a parameterised model was constructed as in G08 and G09. This used Kepler's Laws and assumed a circular orbit to calculate the normalised separation ($z$) of the planet and star centres as a function of time from the stellar mass and radius  ($M_{\star}$ and $R_{\star}$), the planetary mass and radius ($M_{p}$ and $R_{p}$), the orbital period and inclination ($P$ and $i$), and finally a central transit time for each lightcurve ($T_{0,n}$).  The analytic models of \citet{mandel_agol_2002} were then used to calculate the stellar flux occulted by the planet from the normalised separation and the planet/star radius ratio ($\rho$) assuming the quadratic limb darkening function
\[
\frac{I_{\mu}}{I_1} = 1 - a(1-\mu) - b(1-\mu)^2,
\]
where $I$ is the intensity, $\mu$ is the cosine of the angle between the line-of-sight and the normal to the stellar surface, and $a$ and $b$ are the linear and quadratic limb darkening coefficients, respectively.

Limb darkening parameters were obtained from the models of \citet{claret_2000}. We linearly interpolated the ATLAS tables for $T_{eff}$ = 5\,185\,K, log\,$g$ = 4.61, [Fe/H] = 0.27 and $v_{t}$ = 2.0\,kms$^{-1}$ (from T07) to obtain limb darkening parameters in both the $V$ and $R$ bands. The average from the $V$ and $R$ bands was then adopted as our theoretical limb darkening parameters. Several tests were performed to examine the effects of the choice of limb darkening parameters on the results, which are described at the end of this section.

A Markov-Chain Monte-Carlo (MCMC) algorithm was then used to obtain the best fit parameters and their uncertainties \citep[see e.g.,][]{tegmark_2004, holman_winn_2006, cameron_2007, winn_holman_2008}.  This consists of calculating the $\chi^2$ fitting statistic,
\[
\chi^2=\sum_{j=1}^{N}\frac{(f_{j,obs} - f_{j,calc})^2}{\sigma^2_j},  
\]
where  $f_{j,obs}$ is the flux observed at time $j$, $\sigma_j$ is the corresponding uncertainty and  $f_{j,calc}$ is the flux calculated from the model for time $j$ and for the set of physical parameters described above. $M_\star$, $M_p$ and $P$ were held fixed at the values obtained in T07, and we fitted for $i$, $\rho$, $R_\star$ and $T_{0,n}$. The scaling relation $R_{\star} \propto M_{\star}^{1/3}$ was then used to propagate the error in $M_\star$ to $R_\star$ \citep{holman_winn_2006}. Subsequent parameter sets are then chosen by applying small perturbations to the previously accepted set. New parameter sets are always accepted if $\chi^2$ is decreased, or with probability  $\exp(-\Delta\chi^2/2)$ if $\chi^2$ is increased, where $\Delta\chi^2$ represents the difference in $\chi^2$ calculated for the old and new parameter sets. A further two free parameters are added for each light curve to account for the normalisation, namely the out-of-transit flux ($f_{oot,n}$) and a time gradient ($t_{Grad,n}$). The procedure is similar to that used in G08 and G09, to which the reader is referred for details.

Photometric errors $\sigma_j$ were first re-scaled so that the best fitting model for each light curve had a reduced $\chi^2$ of 1. They were then re-scaled once more by a factor $\beta$ to account for correlated noise according to \citet{winn_holman_2008}, again described in G08 and G09. The value determined for $\beta$ depends strongly on the choice of averaging times used to analyse the residuals. Usually the average time in a given range is used. However, in the this analysis we use the maximum value for $\beta$ in the range 10--35 minutes as in G09, to be as conservative as possible in the resulting errors. The values of $\beta$ for each light curve are shown in Table~\ref{tab:RISE_timings}.
 
An initial MCMC analysis was used to estimate the starting parameters and jump functions for $\rho$, $i$, $R_\star$, $T_{0,n}$, $f_{oot,n}$ and $t_{Grad,n}$. An MCMC run was then started for all seven light curves, consisting of five separate chains each with 500\,000 points. The initial parameter set was chosen by adding a $5\sigma$ Gaussian random to each parameters previously determined best fit value, and the first 20\% of each chain was eliminated to keep the initial conditions from influencing the results. The remaining parts of the chains were merged to obtain the best fit values and uncertainties for each free parameter. The best fit value was set as the modal value of the probability distribution, and the $1\sigma$ limits to the values where the integrals of the distribution from the minimum and maximum values were equal to 0.159. To test that the chains had all converged to the same region of parameter space, the Gelman \& Rubin statistic \citep{gelman_rubin_1992} was then calculated for each of the free parameters, and was found to be less than 0.5\% from unity for all parameters, a good sign of mixing and convergence.

To account for errors in the limb darkening parameters, the linear limb darkening coefficient ($a$) was allowed to vary freely whilst holding the quadratic limb darkening coefficient fixed at its theoretical value \citep{southworth_2008}. The chains converged using this technique, and did not require a prior to be set on the limb darkening as required for fitting the TrES-3 light curves in G09. This is likely due to the smaller impact parameter of HAT-P-3 compared to TrES-3. The resulting uncertainties are larger when varying the linear limb darkening parameter, and therefore these values are adopted as our final MCMC solutions.  A further check on the limb darkening parameters was performed by adopting the limb darkening coefficients for the individual $V$ and $R$ filters, and repeating the above procedure, which led to no significant changes in our results.

%________________________________________________________________
\subsection{Central Transit Times}
\label{sect:transit_times}

The central transit times from the MCMC fit described above were adopted as the best fit values. Two different methods were used to account for red noise in the light curves, which is the biggest problem when trying to obtain robust central transit times. The first was re-scaling the error bars by a factor $\beta$ prior to the MCMC fits.

The second method was to use a residual-permutation (RP) or ``prayer bead'' algorithm \citep[see e.g.][]{southworth_2008,gillon_2009}, as described in G09. This consists of reconstructing each light curve from its best fit model and residuals (from the MCMC fit) by adding them, each time shifting the residuals by a random amount. For each reconstructed light curve, a fit is performed by minimising $\chi^2$, and the errors are estimated from the resulting distribution of parameters. Twenty thousand such fits were performed for each light curve, with $M_\star$, $R_\star$, $i$ and $\rho$ selected from a Gaussian distribution at the start of each. The values and uncertainties were taken from T07 for $M_\star$, and from the MCMC fits of the RISE light curves for $i$, $\rho$ and $R_\star$. $T_{0,n}$, $f_{oot,n}$, $T_{Grad,n}$ and $a$ were allowed to vary freely, with their starting points set by randomly selecting a value within 10$\sigma$ from the MCMC best fit values.

For each light curve, the largest uncertainty in the central transit time was selected for the TTV analysis. In all cases this was from the MCMC fitting, which gave errors ranging from $\sim$1.07--1.80 larger than the RP method. This is due to selecting the largest $\beta$ value in a given time range, rather than the average value.

%________________________________________________________________
\section{Results}
\label{sect:results}

%________________________________________________________________
\subsection{System parameters}
\label{sect:results_system_parameters}

The system parameters derived from the MCMC fits of the RISE light curves are given in Table~\ref{tab:system_parameters}. We find $\rho = 0.1098^{+0.0010}_{-0.0012}$, $i = 86.75^{+0.22}_{-0.21}\deg$ and $R_\star = 0.834^{+0.018}_{-0.026} R_\odot$, consistent with the values determined in T07, but with smaller uncertainties. The stellar mass is assumed from T07, and the planet radius and density were determined as $0.890^{+0.021}_{-0.029} R_{Jup}$ and $1.054^{+0.113}_{-0.087}$g\,cm$^{-3}$, respectively. Again, these parameters are consistent with T07, but with smaller uncertainties. 

\begin{table*}
\begin{minipage}{100mm}
\caption{Parameters and $1\sigma$ uncertainties for HAT-P-3 as derived from MCMC fitting of RISE light curves and some further calculated parameters.}
\label{tab:system_parameters}
\begin{tabular}{lccc}
\hline
 Parameter & Symbol & Value & Units\\
\hline
Planet/star radius ratio	&	$\rho$         	& $0.1098^{+0.0010}_{-0.0012}$ 				&\\
Orbital inclination 		&	$i$  			& $86.75^{+0.22}_{-0.21}$		     			& $\deg$\\
Impact parameter		&	$b$  			& $0.576^{+0.022}_{-0.033}$ 					&\\ 
%\\
Transit duration		&	$T_d$		& $2.087^{+0.018}_{-0.014}$ 					& hours\\
Transit epoch			&	$T_0$		& $2454856.70118 \pm 0.00018$ 				&HJD\\
Period				&	$P$			& $2.899738 \pm 0.000007$				& days\\
%\\
Stellar radius$^a$		&	$R_\star$         	& $0.834^{+0.018}_{-0.026}$          & $R_\odot$\\
Planet radius			&	$R_p$         	& $0.890^{+0.021}_{-0.029}$ 					& $R_J$\\
Planet mass$^b$		&	$M_p$		& $0.599 \pm 0.026$  					& $M_J$\\ 
Planet density			&	${\rho}_p$  	& $1.054^{+0.113}_{-0.087}$					& g cm$^{-3}$\\ 
Planetary surface gravity	& 	log $g_p$		& $3.273^{+0.033}_{-0.029}$                                         & [cgs]\\
\hline
\end{tabular}
\medskip
$^a${After error propagated according to $R_\star \propto M_\star^{1/3}$, using $M_\star$ from T07.}
$^b${From T07, displayed here for convenience.}
\end{minipage}
\end{table*}

%________________________________________________________________

\subsection{Transit ephemeris}
\label{sect:transit_ephemeris}

\begin{figure}
\includegraphics[width=84mm]{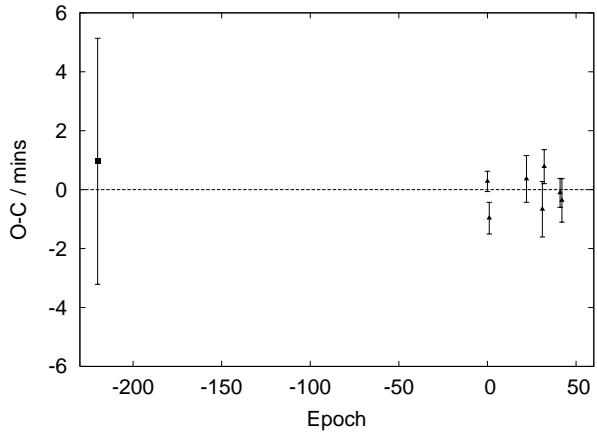}
\caption{
Timing residuals of the RISE transits (triangles) and the discovery epoch from T07 (square) using the updated ephemeris.
}
\label{fig:ttv_plot}
\end{figure}

The central transit times for the RISE light curves are shown in Table~\ref{tab:RISE_timings}. A new ephemeris was calculated by minimising $\chi^2$ through fitting a linear function of Epoch $E$ and Period $P$ to the transit times,
\[
T_c(E) = T_c(0) + EP,
\]
where $E = 0$ was set to the transit from 2009 January 24 taken with RISE, as it has the smallest uncertainty. We included the discovery epoch from T07. The results were $T_c(0) = 2454856.70118 \pm 0.00018$ and $P = 2.899738 \pm 0.000007$. Figure~\ref{fig:ttv_plot} shows a plot of the timing residuals of the transits using this updated ephemeris.

For the data, a straight line fit gives $\chi^2$ = 6.79 for 6 degrees of freedom, or a reduced $\chi^2_{red} = 1.13$. Thus no significant timing signal is found in the timing residuals. Indeed, removing all transits with less than $\sim$20 minutes either before ingress or after egress (Transits $E=$ 1, 22 and 41), as suggested in G09, results in a $\chi^2$ of 2.37 for 5 degrees of freedom ($\chi^2_{red}=0.79$), confirming that a straight line provides a very good fit to the transit times.

\begin{table}
%\begin{minipage}{95mm}
\caption{Central transit times and uncertainties for the RISE photometry including the error source.}
\label{tab:RISE_timings}
\begin{tabular}{lccc}
\hline
Epoch & Central Transit Time & Uncertaintly & $\beta^c$\\
~ & [HJD] & (days) &  ~ \\ 
\hline
0  & 2454856.70137 & 0.00024 & 1.44 \\
1  & 2454859.60024 & 0.00037 & 1.53 \\
22 & 2454920.49567 & 0.00055 & 1.76 \\
31 & 2454946.59260 & 0.00065 & 2.02 \\
32 & 2454949.49334 & 0.00040 & 2.17 \\
41 & 2454975.59037 & 0.00034 & 1.61 \\
42 & 2454978.48993 & 0.00051 & 1.63 \\
\hline
\end{tabular}
\medskip

$^c$ Re-scale factor from red noise analysis (see Section~\ref{sect:system_parameters}).
%\end{minipage}
\end{table}

%________________________________________________________________
\subsection{Limits on a second planet in the HAT-P-3 system.}
\label{sect:upper_mass_limits}

Whilst not showing a significant TTV signal, upper mass limits may still be placed on the presence of a third body in the system as a function of period ratio, by fitting model timing residuals to the timing residuals measured with RISE. The same procedure was used in G09, where the equations of motion were integrated using a 4th-order Runge-Kutta method, with the first two bodies representing the star and planet of the HAT-P-3 system, and the third body representing a hypothetical perturbing planet. The transit times were then extracted when the star and transiting planet align along the direction of observation, and the residuals of a straight line fit are the model timing residuals.

Due to limitations in computation, we assume the amplitude of the timing residuals is proportional to the mass of the perturbing planet \citep{agol_2005,holman_murray_2005}. We also assume that the timing residuals increase with increasing eccentricity of the perturbing planet. Hence, to set upper mass limits we can assume the planets have initially circular  orbits. This is not necessarily true near mean-motion resonance (G09), which is discussed later. The orbits of the planets are assumed to be coplanar throughout.

Models were created for an Earth-mass perturbing planet, with a perturbing planet to transiting planet period ratio distributed from 0.2 to 5.0. The sampling was increased around both the interior and exterior 2:1 resonance (where we expect to probe for the smallest planets), and the transit times were extracted at six directions of observation starting along the y-axis, spaced evenly around the star. Each model was started with the two planets aligned along the y-axis, at opposing sides of the star, and we simulated 2 years of TTVs. The upper mass limit was determined for each model by scaling the mass of the perturbing planet until the $\chi^2$ of the model fit increased by a value $\Delta\chi^2=9$ \citep{steffen_agol_2005,agol_steffen_2007} from that of a linear ephemeris (i.e. a straight line). The $\chi^2$ is then minimised along the epoch, and for a constant added to the residuals, and the mass of the perturbing planet is scaled again until the maximum mass allowed is determined. This is repeated for each observation direction, and the maximum mass found is assumed as our upper mass limit for each period ratio. This process was then repeated twice, with the initial mass of the perturbing planet set to the previously determined upper mass limit. This was found to have little effect on the upper mass limits set, justifying our assumption that the timing residuals scale with the mass of the perturbing planet.

\begin{figure*}
\begin{minipage}{140mm}
\includegraphics[width=\textwidth]{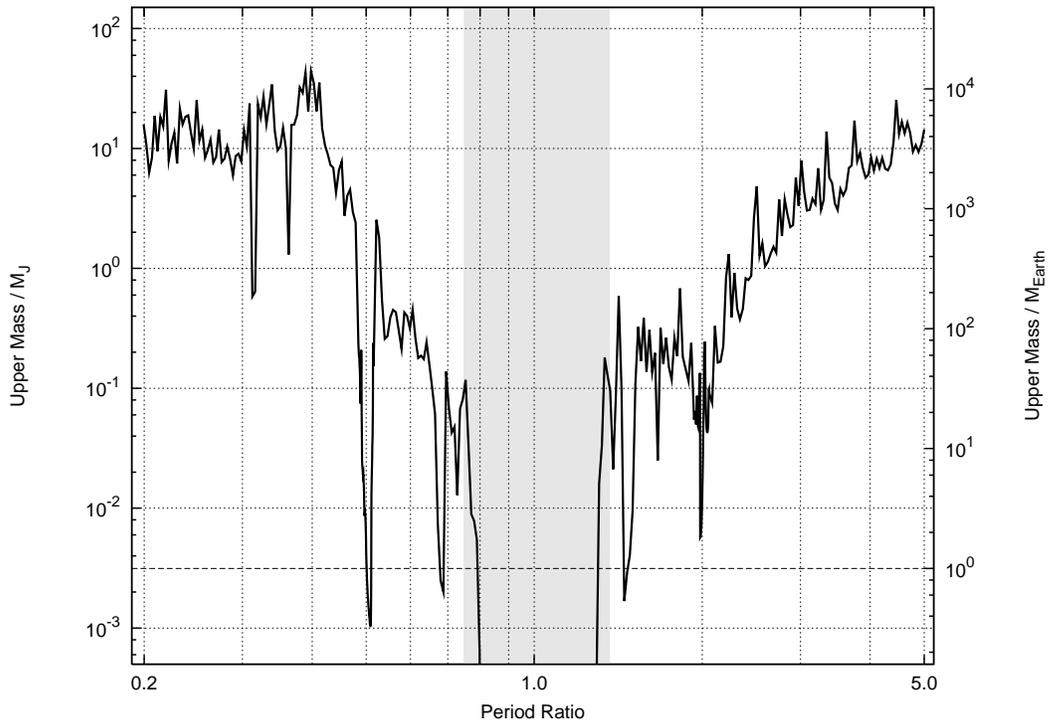}
\caption{
Upper mass limits of a hypothetical 2nd planet in the HAT-P-3 system as a function of period ratio. The solid black line represents the upper mass found for the 3 body simulations, and the horizontal dashed line represents an Earth-mass planet. The region where an Earth-massed planet is not guaranteed to be Hill stable is marked by the grey shading.
}
\label{fig:upper_mass_plot}
\end{minipage}
\end{figure*}

Figure~\ref{fig:upper_mass_plot} shows a plot of the upper mass limits as a function of the period ratio. The solid black line is the upper mass limits set from the three-body simulations, and the horizontal dashed line represents an Earth-mass planet. The results show that our data were sufficiently sensitive to probe for masses as small as 0.33 $M_{\oplus}$ and 1.81 $M_{\oplus}$ in the interior and exterior 2:1 resonance, respectively. The greater sensitivity in the interior 2:1 resonance may be explained in terms of the libration cycle of the timing residuals. The libration period of the interior resonance is $\sim$65 orbits of HAT-P-3b, which allows most of the libration cycle to be covered by the RISE transits. However, the libration period for the exterior 2:1 resonance is longer, $\sim$200 orbits of HAT-P-3b, so only a small portion of the libration cycle is covered by the RISE transits. Therefore, in the exterior 2:1 resonance, lower mass perturbers are not as easily constrained by the transit times.

G09 found the assumption that the smallest timing residuals occur when the perturbing planet has an initially circular orbit is not valid near mean-motion resonance. This was tested in the same way by creating a series of models around the resonances with non-zero eccentricity. A set of models with eccentricities ranging from 0 to 0.23 were created around the exterior 2:1 resonance, and it was found that the amplitude of the timing residuals may drop at low eccentricities by as much as an order of magnitude before rising again at high eccentricities. This may be partly due to the planets getting perturbed by different amounts from their initial period ratios and overshooting the resonance \citep{agol_2005}, resulting in different period ratios when only varying the initial eccentricities.
A similar effect is seen in the interior 2:1 resonance, but with the residuals only dropping by about a factor of $\sim$2. This is a similar conclusion to that in G09, and again suggests that to set true upper mass limits around resonance, the effect of eccentricity needs to be investigated in detail.
These models were then used to place upper mass limits as before, and we found that more realistic upper mass limits would be $\sim$0.7 -- 1.0 $M_{\oplus}$ and $\sim$30 -- 40 $M_{\oplus}$ in the interior and exterior resonance, respectively. Some similar sets of models were produced out of resonance, in which the amplitude of the TTV signal does increase with eccentricity of the perturbing planet, and thus the upper mass limits found are valid.

\citet{barnes_greenberg_2006} explore the stability limits in exoplanet systems, and provide an inequality to test whether a system is Hill stable (equation 2). Using this inequality for the HAT-P-3 system (assuming an Earth-massed perturber), places lower and upper limits on the period ratio of 0.74 and 1.37, respectively. The resulting region not guaranteed to be Hill stable is marked on Figure~\ref{fig:upper_mass_plot} by the grey shading, although stable configurations may still occur in this region. Trojan companions may also exist in stable orbits near the 1:1 resonance. \citet{madhusudan_winn_2009} placed a $2\sigma$ upper mass limit of $\sim260$ Earth masses on a Trojan companion to HAT-P-3b by combining transit observations and radial velocity data.

%________________________________________________________________
\section{Summary and discussion}
\label{sect:summary}

This paper presents the second Transit Timing study using the RISE instrument on the LT, consisting of seven light curves of the exoplanet system HAT-P-3. An MCMC algorithm was used to determine the light curve parameters, which were found to be in agreement with the discovery paper, but with a higher precision.

The central transit times and uncertainties were also determined using the MCMC algorithm, and a residual permutation algorithm was used as an independent check on the errors. Uncertainties in the transit times range from $20-60$ seconds, much greater than the typical transit times obtained for TrES-3 using RISE (G09). This is partly due to the shallower transit of HAT-P-3, but also due to the higher levels of systematic noise in the light curves.

The central transit times were found to be consistent with a linear ephemeris, with $\chi_{red}^2 < 1$ after removing those light curves with less than 20 minutes of data either before ingress or after egress. These transit times were then used to place upper mass limits of an additional planet in the system that could perturb the orbit of HAT-P-3b yet not be detected by our measurements. This showed that we probed for masses as low as  0.33 $M_{\oplus}$ and 1.81 $M_{\oplus}$ in the interior and exterior 2:1 resonance, respectively. However, larger planets may exist in low eccentricity orbits in the 2:1 resonances, and further investigation is required to explore true upper mass limits in a higher dimensional parameter space. These are comparable with the mass limits set using the same techniques for TrES-3 in G09, which had many more light curves available, plus transit times with smaller uncertainties. This is because the period of HAT-P-3 is larger than that of TrES-3, and reflects that the sensitivity of the Transit Timing method increases with the period of the transiting planet.

%________________________________________________________________

\section*{Acknowledgements}

RISE was designed and built with resources made available from Queen's University Belfast, Liverpool John Moores University and the University of Manchester. The Liverpool Telescope is operated on the island of La Palma by Liverpool John Moores University in the Spanish Observatorio del Roque de los Muchachos of the Instituto de Astrofisica de Canarias with financial support from the UK Science and Technology Facilities Council. N.P.G. and E.K.S acknowledge financial support from the Northern Ireland Department for Employment and Learning. D.L.P. was supported by a Leverhulme Research Fellowship for the duration of this work, F.P.K. is grateful to AWE Aldermaston for the award of a William Penney Fellowship, and I.T. acknowledges financial support from the Science and Technology Facilities Council. We thank the anonymous referee, for comments which improved the clarity of this paper.

%________________________________________________________________

%\bibliography{HATP3_bib} % your references Yourfile.bib 
%\bibliographystyle{aa} % style aa.bst

\label{lastpage}

\end{document}